      [1999/12/01 v1.4c Il Nuovo Cimento]
\renewcommand{\thefootnote}{\fnsymbol{footnote}}
\documentclass{cimento}

\title{The interpretation of the {\it Swift} GRB X-ray afterglows}
\author{Yi-Zhong Fan\from{1}\from{2},
} \instlist{
  \inst{1}The Racah Inst. of Physics, Hebrew University, Jerusalem 91904, Israel
  \inst{2}Purple Mountain Observatory, Chinese Academy of
Science, Nanjing 210008, China }

\PACSes{
\PACSit{98.70.Rz}{gamma-ray bursts}}
\begin{document}

\maketitle

\begin{abstract}
We discuss the current interpretations of the {\it Swift} GRB X-ray afterglows,
mainly focusing on the sharp decline at the prompt tail emission, and the shallow
decay afterward, which is then followed by the conventional pre-\emph{Swift} decay
behavior, and the possible X-ray flares during the latter two stages. We emphasize
the role of the central engine in interpreting the GRB afterglows.
\end{abstract}

\begin{figure}
\begin{picture}(0,190)
\put(0,0){\includegraphics{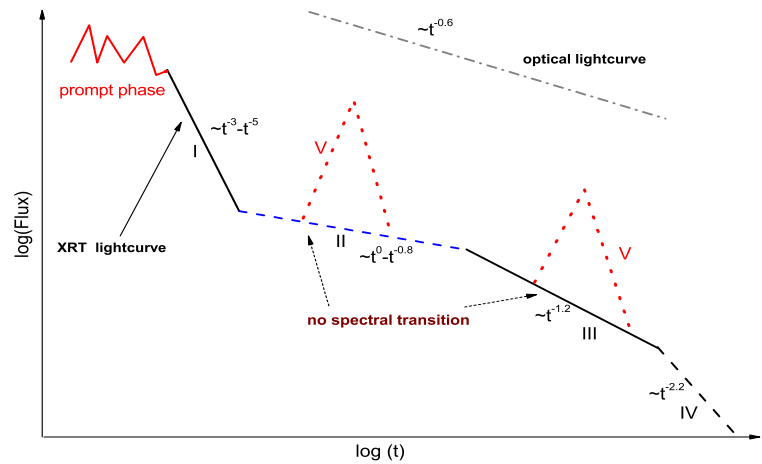}}
\end{picture}
\caption{The schematic cartoon X-ray light curve based on the {\it Swift} XRT data
(see \cite{Zhan06,Nous06} for quite similar plots) as well as  the chromatic break
in optical/X-ray afterglows.} \label{fig:Cartoon}
\end{figure}

Many surprises, mainly in X-ray band, have been brought since the
successful launch of the {\it Swift} satellite in Nov 2004. A
canonical {\it Swift} GRB X-ray afterglow lightcurve has been
summarized by Zhang et al. \cite{Zhan06} and Nousek et al.
\cite{Nous06}. As shown in Figure \ref{fig:Cartoon}, some
interesting features are emerging (detected in a good fraction of
but not all bursts), including the very early sharp decline
preceding the conventional afterglow component (i.e., phase-I), a
shallow decline of the X-ray afterglow before the ``normal" decay
phase (i.e., phase-II), and the energetic X-ray flares (i.e.,
phase-V). In this work, we discuss the interpretation of these
features.

{\bf Phase-I: The rapid decline of the very early X-ray lightcurve.} A steep initial
X-ray decay $F_\nu \propto t^{-\alpha}~(\alpha \sim 3-5)$ has been seen in a good
fraction of {\it Swift} GRBs \cite{Tagl05}, for which the most widely considered
explanation is the off-axis emission from regions at $\theta
>\Gamma^{-1}$ (the curvature effect, or high latitude emission)
\cite{km00}. In the uniform ejecta model, provided that the
sub-outflow powering a prompt $\gamma-$ray pulse is ejected at a
time $t_0$, even after the gamma-rays along the sightline have
ceased (i.e., $t>t_0+\delta t$, where $\delta t$ is the duration
of the pulse), still there is the off-axis emission observed from
$\theta>\Gamma^{-1}$ which follows $ F_{\nu_{\rm x}} \propto
[(t-t_0)/\delta t]^{-2-\beta}$, where $\beta \sim 1$ is the X-ray
spectral index of the emission. Since the central engine turns off
at $t_{\rm turn}$, what we observed is the high latitude emission
of the earlier pulses. The flux declines as $F_{\rm
X}={\sum\limits_{i}} F_{\nu_{\rm x},i} [(t-t_{0,i})/\delta
t_i]^{\rm -(2+\beta_i)}$, where $i$ represents the $i-$th pulse.
Such a decline is much steeper than $(t /t_{\rm
turn})^{-(2+\beta)}$ as long as $t_{\rm turn}\gg \max \{\delta t_i
\}$. What we faced is thus not why {\it phase-I} is so steep but
why it is not significantly steeper unless $ \max \{\delta t_i \}
\sim t_{\rm turn}$, as shown in Figure 2 of \cite{FW05}.

Here we mention two alternatives naturally giving rise to sharp but not too sharp
X-ray decline (i.e., $F_X\propto t^{-3}$ or so). One is the ``dying central engine"
model, in which the central engine does not turn off abruptly. The more and more
dimmer X-ray emission generated in the more and more weaker energy dissipation may
dominate over the curvature effect of the early pulses, resulting in a shallower
decay \cite{FW05}. The other is the ``hot cocoon model". In this model the sharp
decline is explained as due to emission from the hot plasma ``cocoon" associated
with the GRB ejecta, which expands relativistically after the ejecta has broken
through the stellar envelope, if a substantial fraction of the cocoon kinetic energy
is dissipated at scattering optical depths $\sim 10^2-10^3$ \cite{Peer06}.

{\bf Phase-V: The energetic X-ray flares.} So far, energetic X-ray
flares have been detected in several pre-{\it Swift} GRBs and
about half {\it Swift} GRBs \cite{Piro05}. The most widely
considered interpretation is the so-called ``late internal shock"
model, suggested by Fan \& Wei \cite{FW05} and Zhang et al.
\cite{Zhan06}. This model is in light of the following facts: the
very steep decline of the X-ray flares could be interpreted
naturally; the multi flares detected in one burst, as in GRB
050730, could be accounted if the central engine restarts
repeatedly; the simultaneous optical emission is very weak
\cite{Piro05} and may be suppressed by the synchrotron
self-absorption, requiring an energy dissipation radius $<10^{15}$
cm. Considering that the magnetic mechanism may be more efficient
than the neutrino mechanism to extract the energy needed to power
the X-ray flare, Fan et al. \cite{Fan05} suggested that the
outflow might be highly magnetized and the flares were linearly
polarized. In both the ``late internal shock" model and the ``late
internal magnetic dissipation" model, the central engine has to be
able to restart repeatedly. One natural speculation is that such
re-activities are caused by the intermittent fall-back accretion
(see \cite{Zhan06b} for a recent review) \footnote{We propose a
strange star (SS)-white dwarf (WD) merger model for short GRBs
with X-ray flare (Detailed treatment will be presented elsewhere).
We do not discuss a neutron star (NS)-WD merger because the hard
surface of NS acts as a plug, stopping up the accretion. The SS-WD
coalescence is quite similar to a solar mass black hole (BH)
merging with a WD \cite{Frye99}. The short hard $\gamma-$ray
spike(s) is produced when and only when the SS collapses to BH.
The accretion of WD material onto the nascent BH lasts a few
hundred seconds with an accretion rate $\sim$ a few $\times ~
10^{-3}~M_\odot~{\rm s^{-1}}$. The outflow luminosity could be
high to $\sim 10^{48}- 10^{49}~{\rm erg~s^{-1}}$ \cite{Frye99}.
The accretion rate is unsteady, so is the energy output. In this
model, the timescale, the luminosity, and the multipile structure
of the X-ray flare following short GRBs could be well reproduced.
The very late X-ray flares are powered by the fall-back accretion
of the material ejected in the WD-SS merger. There could also be
some precursors---Before the SS collapses to BH, the accretion of
the WD material on the SS may give rise to significant emission.}.

Here we discuss the ``internal refreshed shock" model. It is
speculated that in some bursts, a significant part of the GRB
ejecta energy is carried by the material moving with a Lorentz
factor $\Gamma_{\rm s} \sim 10$ \cite{GK06}. Provided that at a
time $\delta T \sim 100$s after the burst, the central engine
re-starts and launches ultra-relativistic outflow. The newly
launched ejecta (moving with a Lorentz factor $\Gamma_{\rm
inj}\sim $ a few hundred) would catch up with the slow material at
a radius $R_{\rm c}\sim 2\Gamma_{\rm s}^2 c \delta T$ and generate
forward/reverse shock emission. For simplicity, we don't discuss
the detailed process but to assume that {\it surface A} (the
shocked slow material and the shocked new ejecta) has a Lorentz
factor $\Gamma_{\rm acc}$. When the central engine re-activity
turns off, {\it surface A} is at a radius $R_{\rm s} \sim R_{\rm
c}+2\Gamma_{\rm acc}^2 c T_{\rm reac}$, where $T_{\rm reac}$ is
the reactivity timescale of the central engine. The duration of
emission of {\it surface A} (essentially a single pulse) can be
estimated as $T_{\rm A} \sim R_{\rm s}/(2\Gamma_{\rm acc}^2
c)=({\Gamma_{\rm s} \over \Gamma_{\rm acc}})^2 \delta T+ T_{\rm
reac}$.

The observer's timescale $t$ is ``calibrated" to the trigger of the GRB. The forward
shock emission of the GRB ejecta at $R_{\rm c}$ and $\theta=0$ will be observed at
$\sim R_{\rm c}/(2\Gamma^2 c)$. The emission from {\it surface A} at the same point
will reach us at a time $\sim \delta T+T_{90}$, where $T_{90}$ is the duration of
the prompt $\gamma-$ray emission. After the ceasing of the re-activity of the
central engine, the curvature emission component follows
\begin{equation}
F_{\rm A}(t) \propto ({T_{90}+\delta T +t \over T_{\rm
A}})^{-(2+\beta)},
\end{equation}
the decline could be very steep as long as $T_{90}+ \delta T \gg T_{\rm A}$ and thus
could account for some X-ray flare observations.

{\bf Phase-II: The shallow decline of the early X-ray lightcurve and the chromatic
break.} In about half \emph{Swift} GRBs, the X-ray lightcurves are distinguished by
a long term flattening and there is no spectral evolution detected before and after
the shallow-to-``normal" decline transition, which is consistent with a strong
energy injection \cite{Zhan06,Nous06,GK06}. However, for some GRBs with good quality
multi-wavelength afterglow data, the X-ray and optical lightcurves break
chromatically (see Figure \ref{fig:Cartoon} for an illustrative plot) and thus
challenge the energy injection model \cite{FP06b,Pana06b}.

One possible solution is to modify the basic assumption made in
the standard external shock model. Following
\cite{FP06b,Pana06b,Ioka06}, we assume that (i) The density of the
medium is a function of $R$ as $n\propto R^{-k}$; (ii) The shock
physical parameters $\epsilon_e$ and $\epsilon_B$, i.e., the
fractions of shock energy giving to the downstream electrons and
magnetic field, are shock strength dependent; (iii) There might be
an energy injection taking a form $dE_{\rm inj}/dt \propto t^{-q}$
and thus changes the dynamics/emission of the GRB ejecta. The last
two assumptions yield $(\epsilon_e, \epsilon_B)\propto
(\Gamma^{-b},\Gamma^{-c})\propto (t^{{2+q-k \over
2(4-k)}b},t^{{2+q-k \over 2(4-k)}c})\equiv (t^{b'},t^{c'})$ for
$\Gamma\geq \Gamma_0$, otherwise $(\epsilon_e, \epsilon_B)\propto
(\Gamma^{-d},\Gamma^{-e})\propto (t^{{2+q-k \over
2(4-k)}d},t^{{2+q-k \over 2(4-k)}e})\equiv (t^{d'},t^{e'})$, where
$\Gamma_0$ is the Lorentz factor of the outflow at the X-ray
decline transition. The parameters $(b,~c,~d,~e)$ are assumed to
be constant and may be different from each other, so are
$(b',~c',~d',~e')$.

With these three assumptions and following the standard afterglow
treatment \cite{Sari98}, we have the scaling laws (i.e., an
extension of Table 2 of \cite{Zhan06})
\begin{equation}
F_{\nu_\oplus} \propto   \left\{%
\begin{array}{ll}{\epsilon_B (1+Y)^{2/3} t^{(1-q)+{4-6k-2q+3qk\over 3(4-k)}}}
 ~~~~\hbox{for $\nu_\oplus<\nu_c<\nu_m$;} \\
    \epsilon_B^{-1/4}(1+Y)^{-1}t^{(1-q)+{2k-8-qk+4q \over 4(4-k)}}
    ~~~~ \hbox{for $\nu_c<\nu_\oplus <\nu_m$;} \\
    \epsilon_e^{-2/3}\epsilon_B^{1/3}t^{(1-q)+{2+q \over 6}-{k(1-q/2)\over 4-k}}
    ~~~~ \hbox{for $\nu_\oplus <\nu_m<\nu_c$;} \\
    \epsilon_e^{p-1}\epsilon_B^{(p+1)/4} t^{(1-q)-{(p-1)(2+q) \over
4}-{k(1-q/2)\over 4-k}}~~~~\hbox{for $\nu_m<\nu_\oplus <\nu_c$;} \\
   \epsilon_e^{p-1} \epsilon_B^{p-2\over 4}(1+Y)^{-1} t^{(1-q)-{(p-1)(2+q)\over 4}
   +{(2-q)k-8+4q \over 4(4-k)}}~\hbox{for $\nu_\oplus >\max\{ \nu_c, \nu_m \}$,} \\
\end{array}%
\right. \label{eq:Main1}
\end{equation}
where $\nu_\oplus$ is the observer frequency; $\nu_c$ and $\nu_m$
are the standard cooling frequency and the typical synchrotron
radiation frequency of the shocked electrons, respectively; $Y$ is
the synchrotron self-Compton parameter. {\it However, the time
evolving shock parameters render an estimate on $Y$ very
difficult.} For simplicity, one can assume that $Y \sim 0$ when
the synchrotron self-Compton is unimportant and $(1+Y) \sim
\sqrt{\epsilon_e/\epsilon_B}$ when the synchrotron self-Compton is
important (i.e., $Y\gg 1$).

Substituting such a simplified $Y$ into equation (\ref{eq:Main1})
and assuming each set of $(k,~q)$ is fixed (i.e., they take
standard values $(0,~1)$, respectively) or $b'=d'$ and $c'=e'$,
one thus can constrain the parameters $(b',~c',~d',~e')$ or
$(k,~q,~b',~c')$ with the optical and X-ray temporal indexes---the
solution is straightforward since four parameters to be determined
using four relations. But one crucial problem is that such an
approach, in particular the consideration of evolving shock
parameters, is phenomenological and lacks the physical basis.

{\bf Conclusion and Implication.} The GRB afterglow may be attributed to the
continued activity of the central engine and thus should be named as the ``central
engine afterglow" \cite{FPX06}. This idea was first proposed in the context of GRB
970228 \cite{Katz98}. However, the consistency between the predictions of the
afterglow external shock model and most pre-\emph{Swift} afterglow multi-wavelength
observations leads to that this model has been widely accepted. The situation might
change in the \emph{Swift} era. Now it's evident that central engine plays an
important role on shaping the early afterglow. For example, the dying of the central
engine could give rise to the sharp but not very sharp early X-ray decline, i.e.,
{\it phase-I}. The re-activity of the central engine could power the energetic X-ray
flares, i.e., {\it phase-V}. The long activity of the central engine could give rise
to the power-law decaying X-ray afterglow, as found in GRB 060218
\cite{Sode06,FPX06}. We conclude that the role of the central engine should be taken
into account seriously when interpreting GRB afterglows, in contrast to what we
believed before.

\vspace{-0.35cm} \acknowledgments YZF thanks the conference
organizers for partial support and T. Piran, D. M. Wei, B. Zhang
and D. Xu for fruitful cooperations.



\vspace{-0.6cm}
\begin{thebibliography}{0}

\bibitem{Zhan06} \BY{Zhang, B. et al.} \IN{ApJ} {642}{2006}{354}
\bibitem{Nous06} \BY{Nousek, J. A. et al.} \IN{ApJ}{642}{2006}{389}
\bibitem{Tagl05} \BY{Tagliaferri, G. et al.} \IN{Nature} {436}
{2005} {985}; \BY{Goad, M. et al.} \IN{A\&A} {449} {2006} {89}
\bibitem{km00} \BY{Kumar, P. and Panaitescu, A.} \IN{ApJ} {541} {2000}
{L51}
\bibitem{FW05} \BY{Fan, Y. Z. and Wei, D. M.} \IN{MNRAS}{364}{2005}{L42}
\bibitem{Peer06} \BY{Pe'er, A., M\'esz\'aros, P. and Rees, M. J.}
\IN{ApJ} {642} {2006} {995}
\bibitem{Piro05} \BY{Piro, L. et al.} \IN{ApJ} {623}{2005}{314};
 \BY{Burrows, D. N. et al.} \IN{Science}{309}{2005}{1833}
\bibitem{Fan05} \BY{Fan, Y. Z., Zhang, B. and Proga, D.} \IN{ApJ} {635} {2005} {L129}
\bibitem{Zhan06b} \BY{Zhang, B.} \IN{AIPC} {838} {2006} {392}
\bibitem{Frye99} \BY{Fryer C. L., Woosley S. E., Herant M. and Davis M.
B.} \IN{ApJ} {520} {1999} {650}
\bibitem{GK06} \BY{Granot, J. and Kumar, P.} \IN{MNRAS} {366} {2006}
{L13}
\bibitem{FP06b} \BY{Fan, Y. Z. and Piran, T.} \IN{MNRAS}{369}{2006}{197}
\bibitem{Pana06b} \BY{Panaitescu, A. et al.} \IN{MNRAS}
{369}{2006}{2059}
\bibitem{Ioka06} \BY{Ioka, K., Toma, K., Yamazaki, R. and Nakamura, T.}
\IN{A\&A}{458}{2006} {7}
\bibitem{Sari98} \BY{Sari, R., Piran, T. and Narayan, R.}
\IN{ApJ}{497}{1998}{L17}; \BY{Chevalier R. A. and Li Z. Y.}
\IN{ApJ}{536}{2000}{195}
\bibitem{FPX06} \BY{Fan, Y. Z., Piran, T. and Xu, D.}
\IN{JCAP}{0609}{2006}{013}
\bibitem{Katz98} \BY{Katz, J. I., Piran, T. and Sari, R.} \IN{Phys. Rev.
Lett.} {80}{1998}{1580}
\bibitem{Sode06} \BY{Soderberg, A.  et al.} \IN{Nature}
{442}{2006}{1014}

\end{thebibliography}
\end{document}